\documentclass[preprint,12pt]{elsarticle}
\usepackage{epsfig}%
\usepackage{amsmath}%
\usepackage{graphicx}
\usepackage{amsmath}%
\usepackage{amsfonts}%
\usepackage{amssymb}

\begin{document}
\begin{frontmatter}
\title{Geometrical scaling for identified particles}
\author{Michal Praszalowicz\fnref{fn1}}
\address{M. Smoluchowski Institute of Physics, Jagellonian University, \\
Reymonta 4, 30-059 Krakow, Poland}
 \fntext[fn1]{e-mail: {\tt michal@if.uj.edu.pl}}

\begin{abstract}
We show that recently measured transverse momentum spectra of identified particles
exhibit geometrical scaling (GS) in scaling variable 
$\tau_{\tilde{m}_{\rm T}}=(\tilde{m}_{\rm T}/Q_0)^{2} (\tilde{m}_{\rm T}/ W)^{\lambda}$
where $\tilde{m}_{\rm T}=\sqrt{m^2+p^2_{\rm T}}-m$. We explore consequences of GS
and show that both mid rapidity multiplicity and mean transverse momenta grow as powers
of scattering energy. Furthermore, assuming Tsallis-like parametrization of the spectra
we calculate the coefficients of this growth. We also show that Tsallis temperature is
related to the average saturation scale.
\end{abstract}

\end{frontmatter}

\setcounter{footnote}{0}

\section{Introduction}

Geometrical scaling (GS) \cite{Stasto:2000er} has been observed both in Deep
Inelastic Scattering (DIS) at HERA
\cite{GolecBiernat:1998js,Praszalowicz:2012zh} and in particle production in
hadronic collisions
\cite{McLerran:2010ex,Praszalowicz:2011tc,Praszalowicz:2011rm}. It is an
immediate consequence of the existence of an intermediate energy scale called
saturation scale, denoted hereafter as $Q_{\mathrm{s}}$. Saturation
\cite{sat1,sat2} (for a review see \cite{Mueller:2001fv,McLerran:2010ub})
appears due to the nonlinearities of parton evolution at small Bjorken $x$.
General form of this evolution is given by the so-called JIMWLK equations
\cite{jimwlk} which in the large $N_{\mathrm{c}}$ limit reduce to the
Balitsky-Kovchegov equation \cite{BK}. These equations have 
traveling wave solutions which explicitly exhibit GS \cite{Munier:2003vc}. An
effective theory relevant for the small Bjorken $x$ region is so called Color
Glass Condensate (CGC) \cite{MLV}. For the purpose of present work the details
of the saturation are not of primary importance; it is the very existence of the
saturation scale which plays the crucial role.

For processes with an external scale ($Q^{2}$ in DIS or $p_{\mathrm{T}}^{2}$
of an observed particle produced in hadron-hadron scattering) larger than a
typical non-perturbative energy scale $\Lambda^{2}$ being of the order of a
couple of hundreds MeV, and smaller than, say, 10 GeV where perturbative QCD
can be applied, observables like photon-proton cross-section or charged
particle multiplicity depend only upon the ratio of this scale to
$Q^{2}_{\mathrm{s}}$ called scaling variable $\tau$. This property is referred
to as geometrical scaling \cite{Stasto:2000er}. However, in a situation where
two (or more) external energy scales are present, there exist more than one
such ratios, what implies violation or at least modification of GS. Indeed, as
we have shown in Ref.~\cite{Praszalowicz:2013uu}, particle production in
forward rapidity region provides a \emph{bona fide} example of GS violation.
Another possible case of interest are spectra of identified particles where
particle masses provide yet another external energy scale that might lead to
GS violation \cite{Stebel:2013kca}. It is the purpose of the present paper to
see whether this is really the case.

It is well known that particle spectra at low and medium transverse momenta
can be described by thermal distributions in transverse mass $m_{\mathrm{T}%
}=\sqrt{p^{2}_{\mathrm{T}}+m^{2}}$ with "temperature" $T$ which is a function
of the scattering energy \cite{Hagedorn}. One may therefore hope that in some
limited range geometrical scaling can be still present with pertinent scaling
variable being $m^{2}_{\mathrm{T}}/Q^{2}_{\mathrm{s}}$. It is also known that
more accurate fits are obtained by means of Tsallis-like parametrization
\cite{Tsallis} where particle multiplicity distribution takes the following
form (see \emph{e.g.} \cite{Chatrchyan:2012qb}):
\begin{equation}
\frac{1}{p_{\text{T}}}\frac{d^{2}N}{dydp_{\text{T}}}=C\frac{dN}{dy}\left[
1+\frac{m_{\mathrm{T}}-m}{n\,T}\right]  ^{-n}\label{Tsallis}%
\end{equation}
with%
\begin{equation}
C=\frac{(n-1)(n-2)}{n\,T\left(  n\,T+(n-2)m\right)  }.\label{normC}%
\end{equation}
Coefficient $C$ in Eq.~(\ref{normC}) ensures proper normalization of
(\ref{Tsallis}). Here $n$ and $T$ are free fit parameters that depend on
particle species. In the limit $n \rightarrow\infty$ distribution
(\ref{Tsallis}) tends to the exponent $\exp(-m_{\mathrm{T}}/T)$, \emph{i.e.}
to the thermal distribution mentioned above. Formula (\ref{Tsallis}) suggests
yet another possibility of scaling variable, namely $\tilde{m}^{2}%
_{\mathrm{T}}/Q^{2}_{\mathrm{s}}$ where
\begin{equation}
\tilde{m}_{\mathrm{T}}=m_{\mathrm{T}}-m=\sqrt{p^{2}_{\mathrm{T}}+m^{2}}-m
.\label{tildemT}%
\end{equation}
Subtracting $m$ from $m_{\mathrm{T}}$ in (\ref{tildemT}) contributes in the
large $n$ limit to an overall factor and does not influence the functional
dependence of thermal distribution. For finite $n$, as we shall see, it has a
significant impact on the shape of the multiplicity distribution. Moreover,
variable $\tilde{m}_{\mathrm{T}}$ is "more similar" to $p_{\mathrm{T}}$ as it
vanishes for $p_{\mathrm{T}} \rightarrow0$ for all particle species, while
$m_{\mathrm{T}}$ goes to the species dependent threshold value $m_{\alpha}$. For large
transverse momenta both $\tilde{m}_{\mathrm{T}}$ and $m_{\mathrm{T}}$ tend to
$p_{\mathrm{T}}$.

Saturation scale $Q_{\mathrm{s}}$ characterizes small Bjorken $x$ gluon cloud
developed in the colliding hadrons due to the BFKL-like evolution. It
depends upon gluons' $x$'s which for elastic scattering of massless gluons are
given by
\begin{equation}
x=\frac{p_{\mathrm{T}}}{\sqrt{s}}\,e^{\pm y}\label{Bjx}%
\end{equation}
where $W=\sqrt{s}$ denotes c.m.s. scattering energy, $p_{\mathrm{T}}$ and $\pm
y$ refer to the transverse momentum and to the rapidities of scattered gluons.
Saturation scale used originally in
Refs.~\cite{Stasto:2000er,GolecBiernat:1998js} takes the following form
\begin{equation}
Q_{\mathrm{s}}^{2} = Q_{0}^{2} \left(  \frac{x}{x_{0}}\right) ^{-\lambda
}\label{Qsdef}%
\end{equation}
where $x_{0}$ is of the order of $10^{-3} - 10^{-4}$. Throughout this paper we
shall assume $x_{0} =10^{-3} $. Our results, however, are not sensitive
neither to this choice nor to the value of $Q_{0}$ for which we take 1~GeV/$c$. It
follows from Eq.~(\ref{Bjx}) that in the case of (unidentified) charged
particles spectra, scaling variable is naturally given as
\cite{McLerran:2010ex}
\begin{equation}
\tau_{p_{\rm T}} =\frac{p_{\text{T}}^{2}}{Q_{\mathrm{s}}^{2}}=\frac{p_{\text{T}}^{2}%
}{Q_{0}^{2}} \left(  \frac{p_{\text{T}}}{W}\right)  ^{\lambda}.\label{taupdef}%
\end{equation}
Due to our choice of $x_{0}$, transverse momentum in Eq.~(\ref{taupdef}) is
given in GeV/$c$ and scattering energy $W$ in TeV. This form of scaling variable
has been successfully tested against data for p-p collisions at the LHC
\cite{McLerran:2010ex, Praszalowicz:2011tc,Praszalowicz:2011rm} and also at
lower energies of NA61-SHINE experiment at CERN SPS \cite{Praszalowicz:2013uu}, 
as well as in the case of heavy ion collisions at RHIC
\cite{Praszalowicz:2011rm}.

Here we propose that in the case of identified particles another scaling
variable should be used in
which $p_{\mathrm{T}}$ is replaced by $\tilde{m}_{\mathrm{T}}$ ($\tilde
{m}_{\mathrm{T}}$ -- scaling), \emph{i.e.}
\begin{equation}
\tau_{\tilde{m}_{\rm T}} =\frac{\tilde{m}_{\text{T}}^{2}}{Q_{0}^{2}}\left(
\frac{\tilde{m}_{\text{T}}}{W}\right)  ^{\lambda}.\label{taumtdef}%
\end{equation}
This choice is purely phenomenological for the following reasons. Firstly, the
gluon cloud is in principle not sensitive to the mass of the particle it
finally is fragmenting to, so in principle one should take $p_{\mathrm{T}}$ as
an argument of the saturation scale. In this case the proper scaling variable
would be
\begin{equation}
\tau_{\tilde{m}_{\rm T} p_{\rm T}} =\frac{\tilde{m}_{\text{T}}^{2}}{Q_{0}^{2}}\left(
\frac{p_{\text{T}}}{W}\right)  ^{\lambda}.\label{taumtpdef}%
\end{equation}
We shall show, however, that this choice ($\tilde{m}_{\mathrm{T}}%
$$p_{\mathrm{T}}$ -- scaling) does not really differ numerically from the one
given by Eq.~(\ref{taumtdef}). On the other hand Eq.~(\ref{taumtdef}) has an
advantage over (\ref{taumtpdef}) since, as we shall see, it allows to
calculate analytically many properties of the spectra assuming Tsallis form of
the scaling function (\ref{Tsallis}).

Secondly, if we would take seriously kinematics for massive particle
production, then Bjorken $x$ would be given by Eq.~(\ref{Bjx}) with
$p_{\mathrm{T}}$ replaced by $m_{\mathrm{T}}$. Hence the natural choice for
the scaling variable would be ($m_{\mathrm{T}}$ -- scaling)
\begin{equation}
\tau_{{m}_{\text{T}}} =\frac{{m}_{\text{T}}^{2}}{Q_{0}^{2}}\left(  \frac{m_{\text{T}}}%
{W}\right)  ^{\lambda}.\label{taumdef}%
\end{equation}
We shall show, however, that for this choice of the scaling variable GS is not present.

The paper is organized as follows. In Sect.~\ref{scaling} we shall test GS
scaling hypothesis on the recent ALICE data for identified particles
\cite{ALICE}. We will see that $p_{\mathrm{T}}$ spectra exhibit $\tilde
{m}_{\mathrm{T}}$--scaling in variable (\ref{taumtdef}). After that, in
Sect.~\ref{conseq}, we shall examine the consequences of GS as far as the energy
dependence of total multiplicity and mean transverse momentum is concerned. We
shall see that power-like growth of both of them is a natural consequence of
GS. Finally in Sect.~\ref{shape} we shall match hypothesis of GS and
phenomenological observation that particle spectra are well described by the
Tsallis-like distribution. It will be shown that Tsallis temperature is
proportional to the average saturation scale depending only on the scattering
energy $W$, whereas Tsallis exponent $n$ is in the first approximation energy
independent. Finally, we shall conclude in Sect.~\ref{concl}.

%%%%%%%%%%%%%%%%%%%%%%%%%%%%%%%%%%%%%%%%

\section{Scaling properties of $p_{\mathrm{T}}$ distributions of identified
particles}

\label{scaling}

Throughout this paper we shall use recent ALICE data for the $p_{\mathrm{T}}$
spectra at 0.9, 2.76 and 7~TeV \cite{ALICE}. In the latter case the data cover wide
$p_{\mathrm{T}}$ range from 0.1 to 19 GeV/$c$ (for pions). Unfortunately
available pion data for 0.9~TeV span over much narrower range: 0.1 - 2.5
GeV/$c$, and for 2.76~TeV from 2.1 to 19 GeV/$c$, and similarly for kaons and
protons where, however, there is no overlap between 0.9~TeV and 2.76~TeV
points. For this reason the analysis presented in this paper can be only qualitative.

\begin{figure}[h!]
\centering
\includegraphics[scale=0.33]{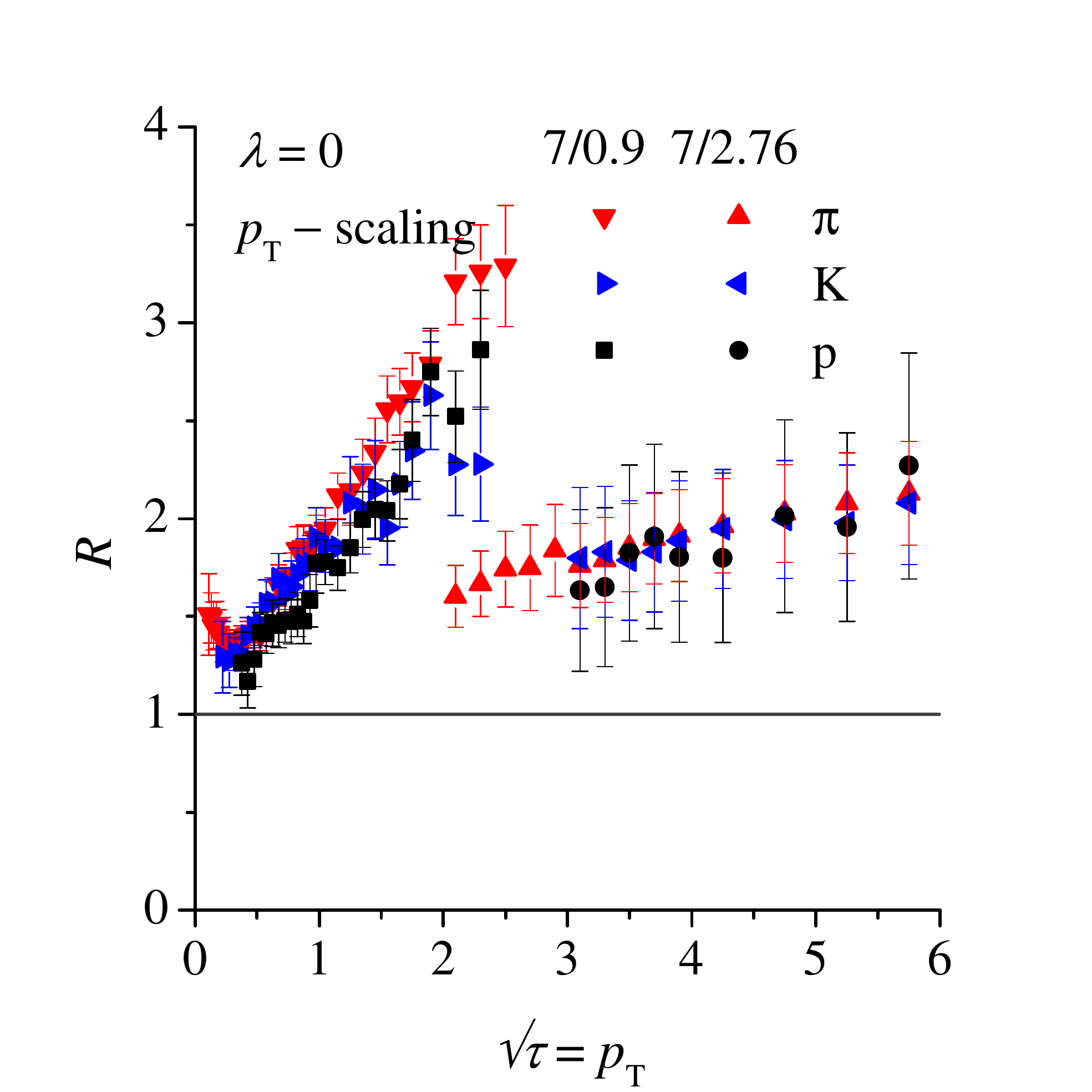}
\includegraphics[scale=0.33]{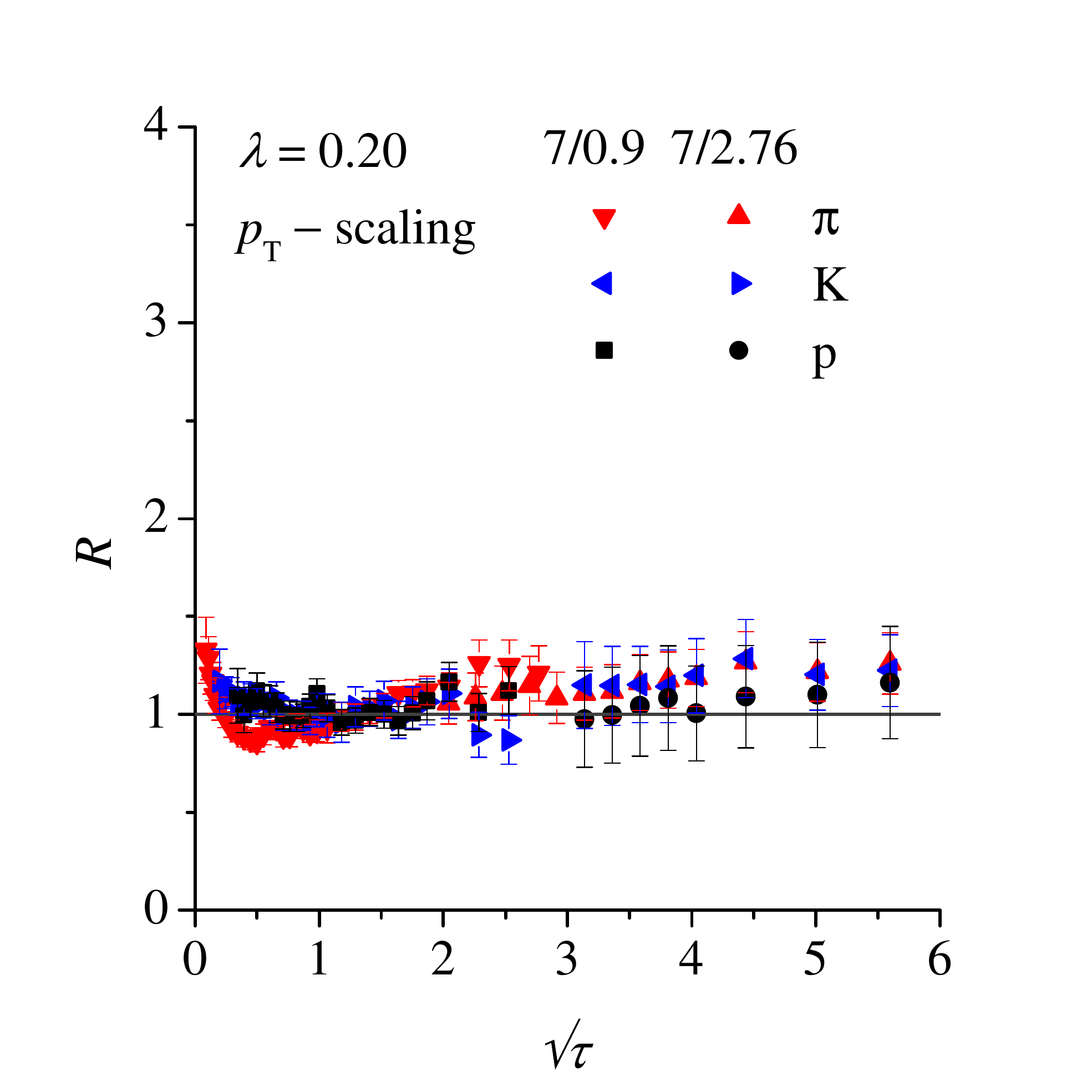}\\
\includegraphics[scale=0.33]{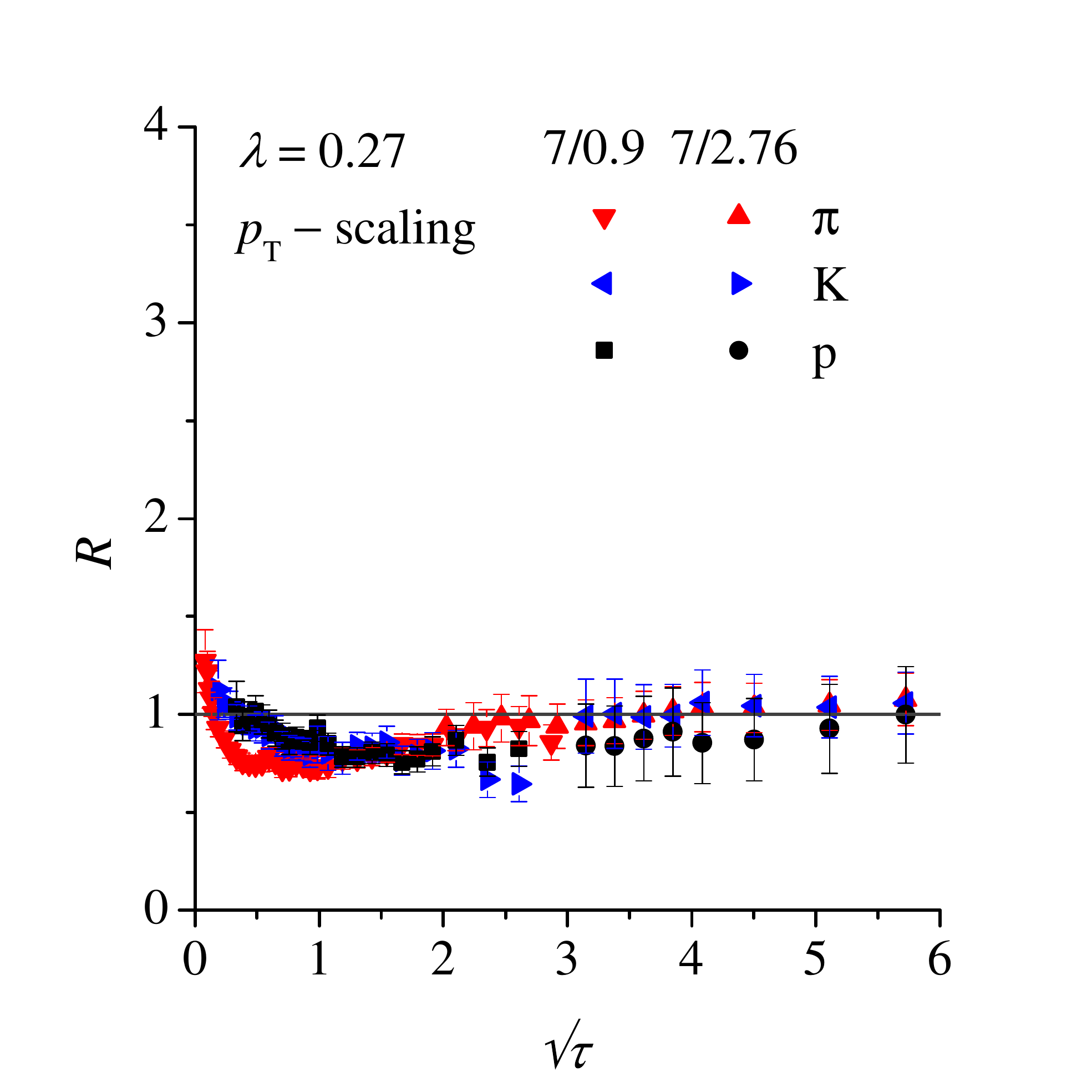}
\includegraphics[scale=0.33]{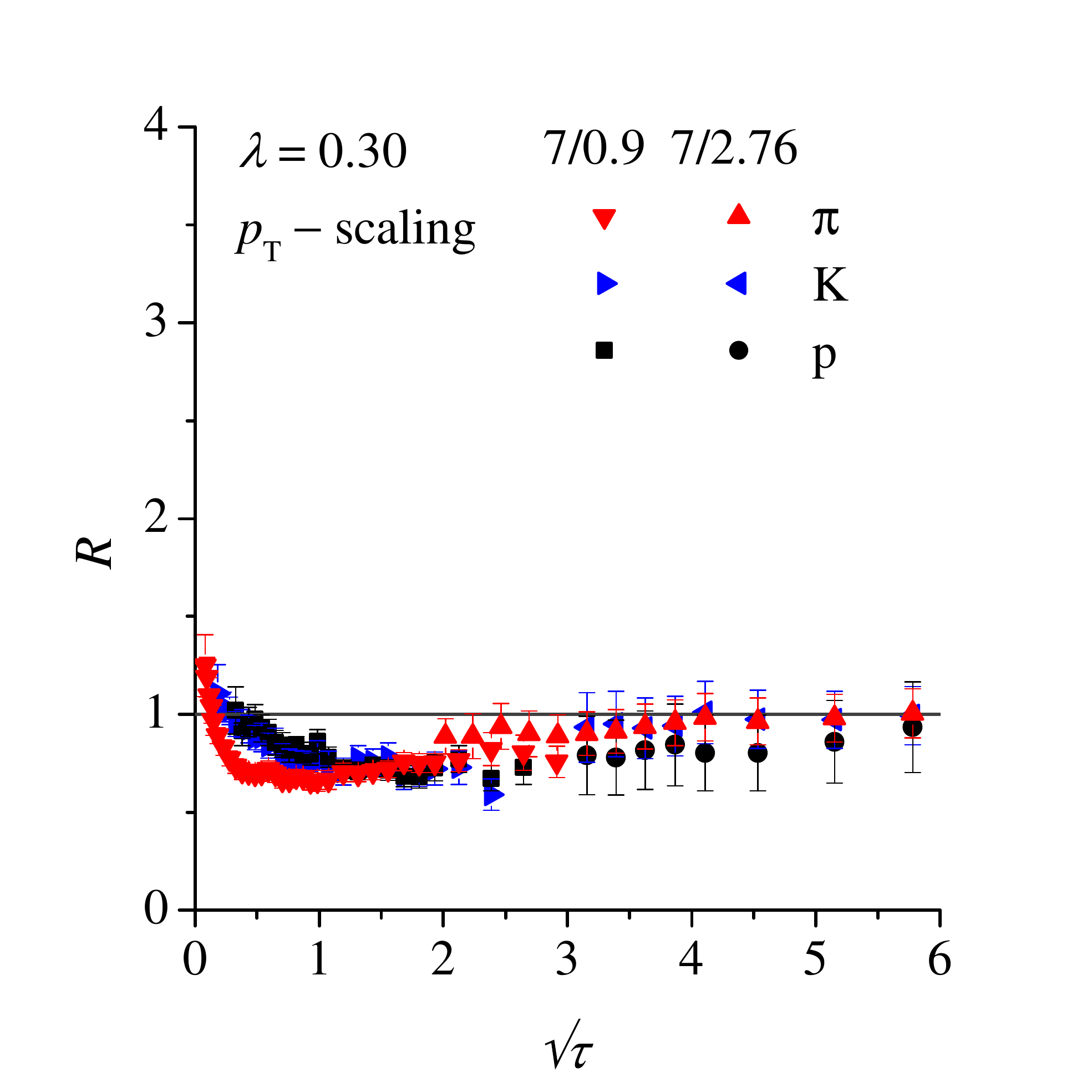}
\caption{Illustration of
geometrical scaling in scaling variable $\tau_{p_{\rm T}}$. Multiplicity ratios
$R_{W_{1}/W_{2}}$ for $W_{1}=7$~TeV are plotted as functions of scaling
variable $\tau_{p_{\rm T}}$ for pions (red triangles: "up" for $W_{2} =2.76$~TeV,
"down" for $W_{2}=0.9$~TeV) kaons (blue triangles: "right" for $W_{2}
=2.76$~TeV, "left" for $W_{2}=0.9$~TeV) and protons (back circles for $W_{2}
=2.76$~TeV and black squares $W_{2}=0.9$~TeV) for different values of the
exponent $\lambda$ a) $\lambda=0$, b) $\lambda=0.2$, c) $\lambda=0.27$ and d)
$\lambda=0.3$.}%
\label{fig:rpT}
\end{figure}

\newpage

In order to asses the quality of GS we shall apply the \emph{method of ratios}
used previously in Refs.~\cite{Praszalowicz:2011rm,Praszalowicz:2013uu} 
in the context of hadron scattering and in 
Refs.~\cite{Praszalowicz:2012zh} for DIS.
Hypothesis of GS means that particle spectra measured at different energies
$W$ are equal when expressed in terms of scaling variable $\tau$
(\ref{taupdef}) -- (\ref{taumtpdef}) or (\ref{taumdef}).
Therefore for each particle species $\alpha$ we have
\begin{equation}
\left.  \frac{1}{p_{\text{T}}}\frac{d^{2}N_{\alpha}}{dy dp_{\text{T}}}\right\vert
_{\left\vert y\right\vert <y_{0}}=\frac{1}{Q_{0}^{2}}F_{\alpha}(\tau
)\label{GS}%
\end{equation}
where $F_{\alpha}(\tau)$ is energy independent function of scaling variable
$\tau$ which, however, may depend on particle species $\alpha$. Here $y_{0}$
is the rapidity cut (assumed to be small) which we shall omit in the
following. Therefore, if hypothesis of GS is true, we expect that the ratios
of multiplicity distributions at two different energies $W_{1}$ and $W_{2}$
(denoted hereafter as $R_{W_{1}/W_{2}}$) should be equal to unity if expressed
in terms of scaling variable $\tau$. For the purpose of the present analysis
we chose $W_{1}=7$~TeV as the reference energy.

In Fig.~\ref{fig:rpT}
we plot ratios $R_{7/0.9}$ and $R_{7/2.76}$ as functions of scaling variable
$\sqrt{\tau}=\sqrt{\tau_{p_{\rm T}}}$ (\ref{taupdef}) for different choices of
exponent $\lambda$ entering the definition of the saturation scale
$Q_{\mathrm{s}}$ (\ref{Qsdef}). We see that for $\lambda=0$ when $\sqrt
{\tau_{p_{\rm T}}}=p_{\mathrm{T}}/Q_{0}$ ratios are substantially larger than 1 and
grow with $p_{\mathrm{T}}$. When $\lambda$ is increased the ratios get smaller
and flatter. One can see that the optimal value of exponent $\lambda$ is
somewhere between 0.2 and 0.27. This value is a bit smaller than the value
$\lambda=0.27$ obtained in the analysis of unidentified NSD spectra measured
by the CMS collaboration at the LHC \cite{McLerran:2010ex}. 
We can see that there is a dip in these
ratios around $\sqrt{\tau} \sim 1$ which is especially pronounced
for pions. Finally, let us remark that we plot $R_{W_{1}/W_{2}}$ only up to
$\sqrt{\tau}=6$; we shall see that for larger values of $\sqrt{\tau}$ the
ratios get larger than 1 and start growing with $\sqrt{\tau}$. Hence we
conclude that there is a window of GS delimited from below by nonperturbative
physics and from above by perturbative production of particles with high
transverse momentum \cite{Praszalowicz:2013uu}.

\begin{figure}[h!]
\centering
\includegraphics[scale=0.33]{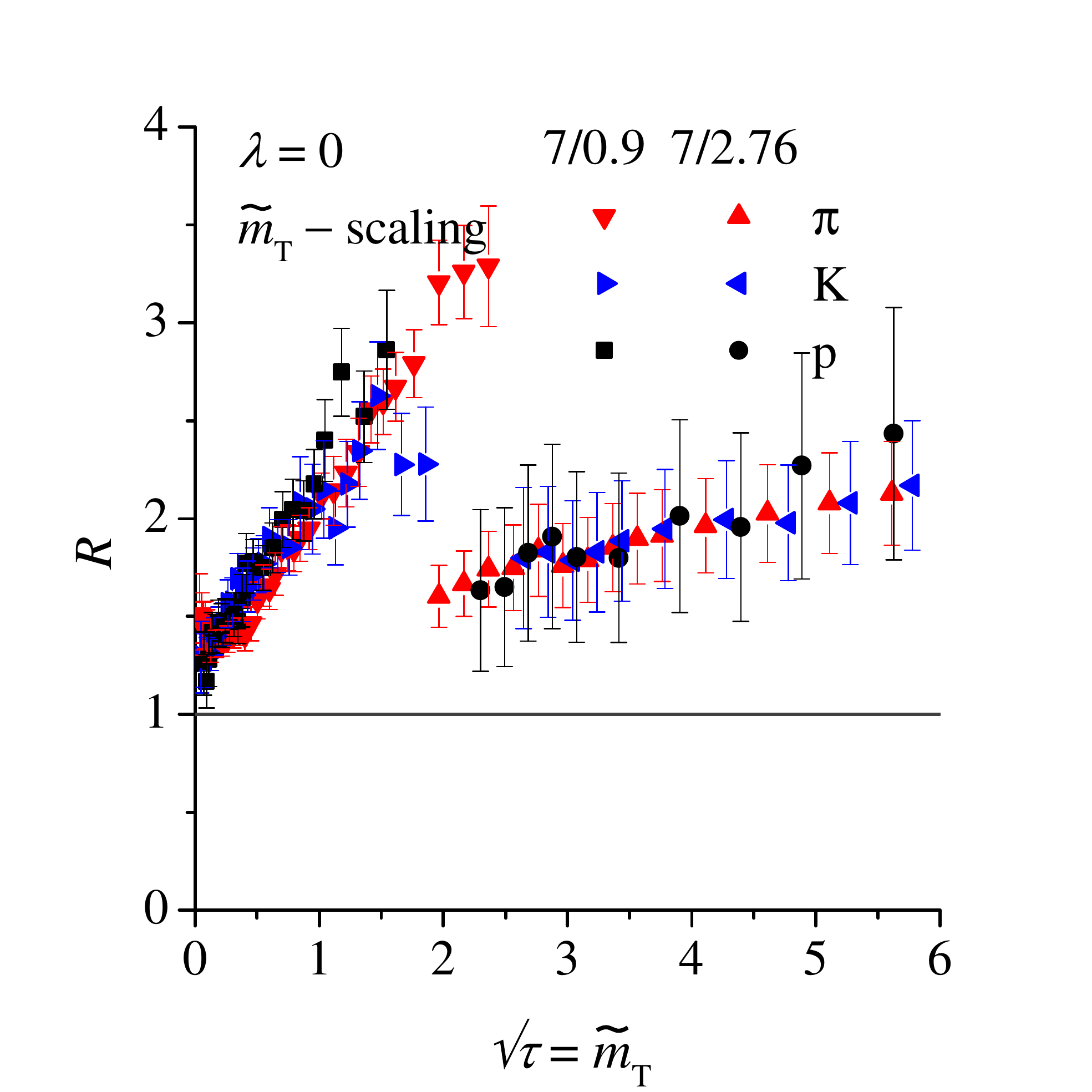}
\includegraphics[scale=0.33]{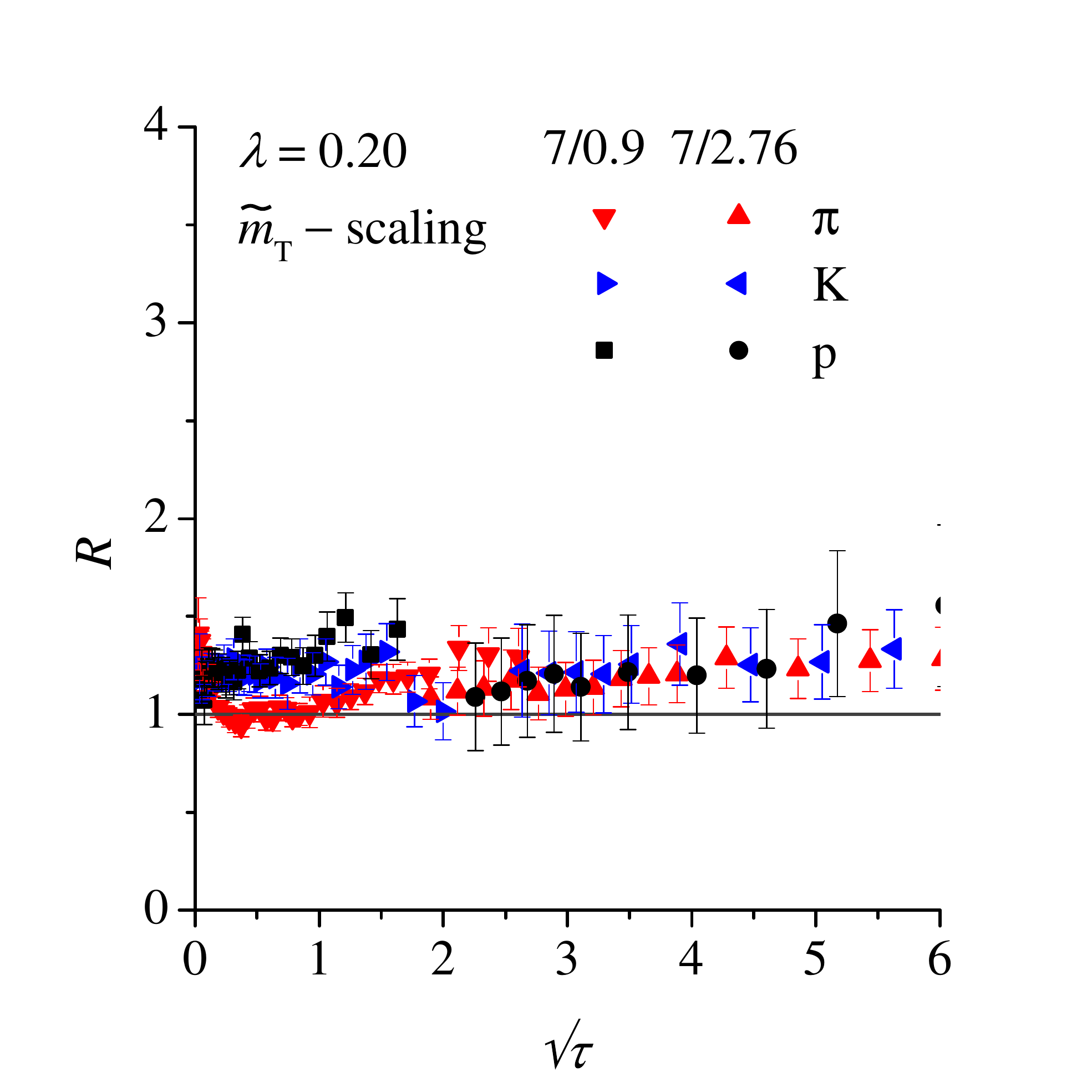}\\
\includegraphics[scale=0.33]{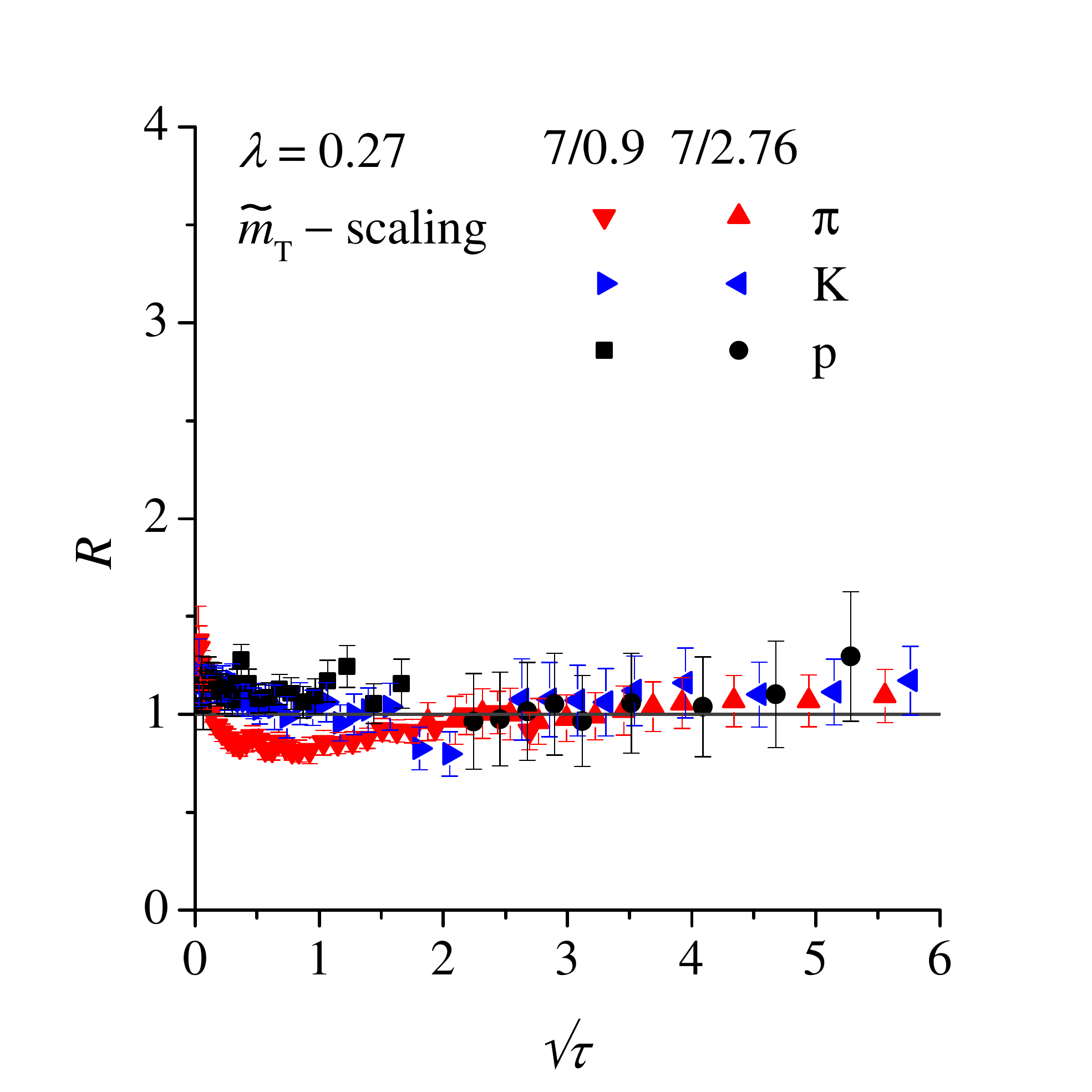}
\includegraphics[scale=0.33]{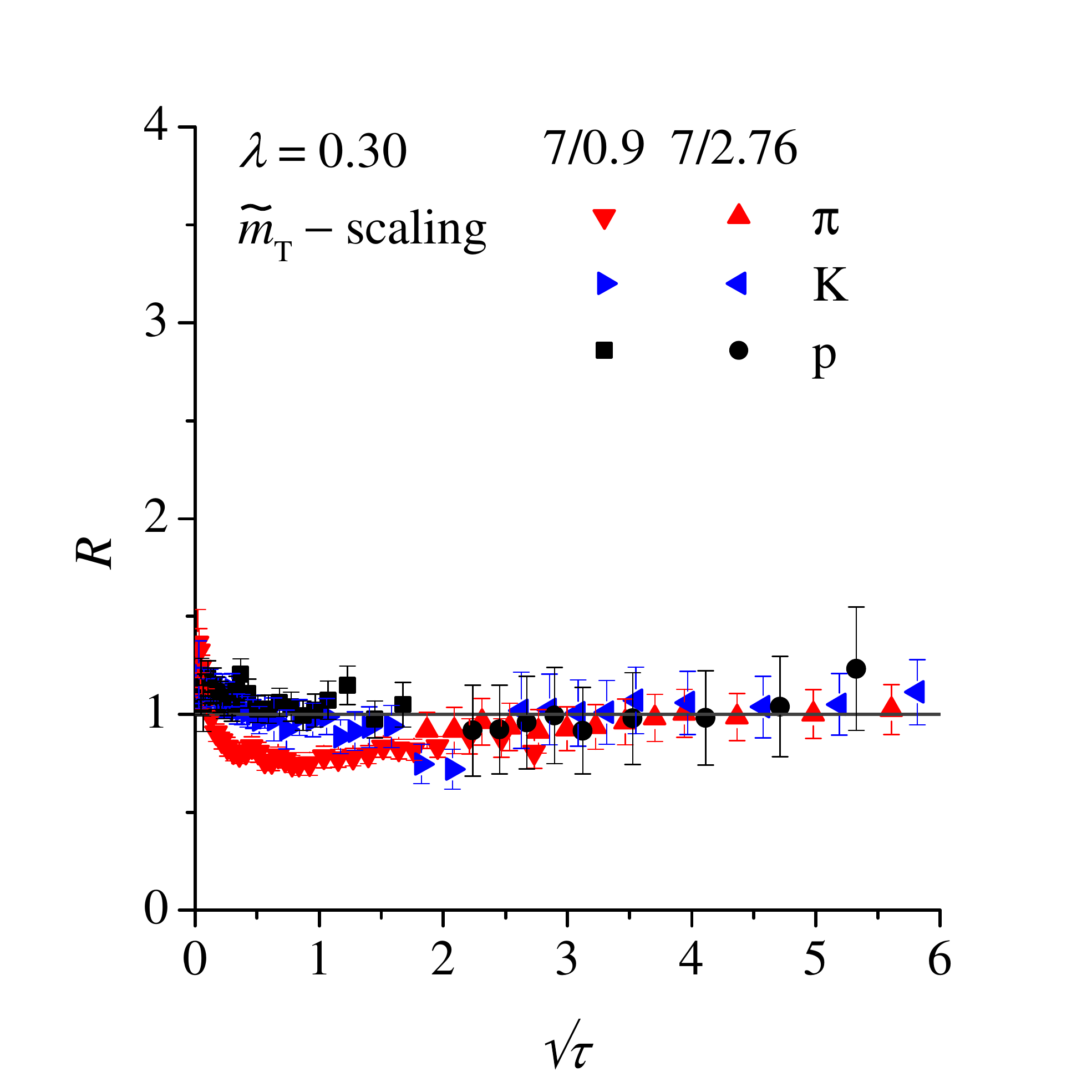}
\caption{Illustration of
geometrical scaling in scaling variable $\tau_{\tilde{m}_{\rm T}}$. Multiplicity
ratios $R_{W_{1}/W_{2}}$ for $W_{1}=7$~TeV are plotted as functions of scaling
variable $\tau_{\tilde{m_{\rm T}}}$ for pions (red triangles: "up" for $W_{2}
=2.76$~TeV, "down" for $W_{2}=0.9$~TeV) kaons (blue triangles: "right" for
$W_{2} =2.76$~TeV, "left" for $W_{2}=0.9$~TeV) and protons (back circles for
$W_{2} =2.76$~TeV and black squares $W_{2}=0.9$~TeV) for different values of
the exponent $\lambda$ a) $\lambda=0$, b) $\lambda=0.2$, c) $\lambda=0.27$ and
d) $\lambda=0.3$.}%
\label{ratiosmT}%
\end{figure}

Although one can conclude from Fig.~1 that GS works reasonably well for
standard scaling variable $\tau_{p_{\rm T}}$ (\ref{taupdef}), we are going to examine
now the hypothesis that the proper scaling variable for identified particles
is $\tau_{\tilde{m}_{\rm T}}$ of Eq.~(\ref{taumtdef}). To this end in
Fig.~\ref{ratiosmT} we plot again ratios $R_{7/0.9}$ and $R_{7/2.76}$ for four
different choices of exponent $\lambda$ as functions of scaling variable
$\sqrt{\tau_{\tilde{m}_{\rm T}}}$. We see that the dip for small values of  $\tau$ has
basically disappeared for kaons and protons and has been largely reduced for
pions. Moreover, good quality GS scaling has been achieved for larger value of
exponent $\lambda\sim 0.27 - 0.3$ in fair agreement with analysis of DIS \cite{Praszalowicz:2012zh}. 
In order to
quantify this statement one has to wait, however, until lower energy data is
published for larger $p_{\mathrm{T}}$ range similar to the one at $W_{1}=7$~TeV.

\newpage

\begin{figure}[h!]
\centering
\includegraphics[scale=0.33]{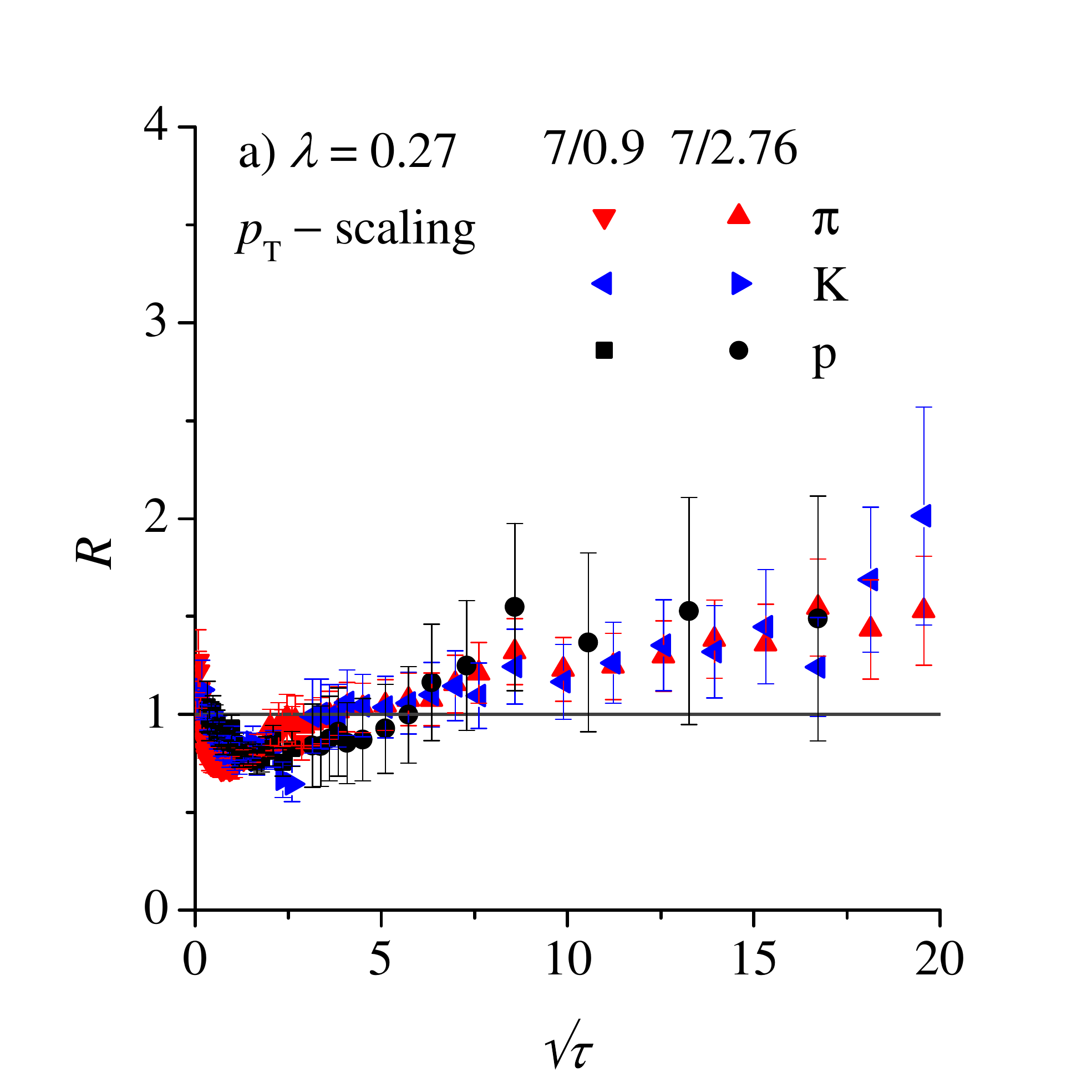}
\includegraphics[scale=0.33]{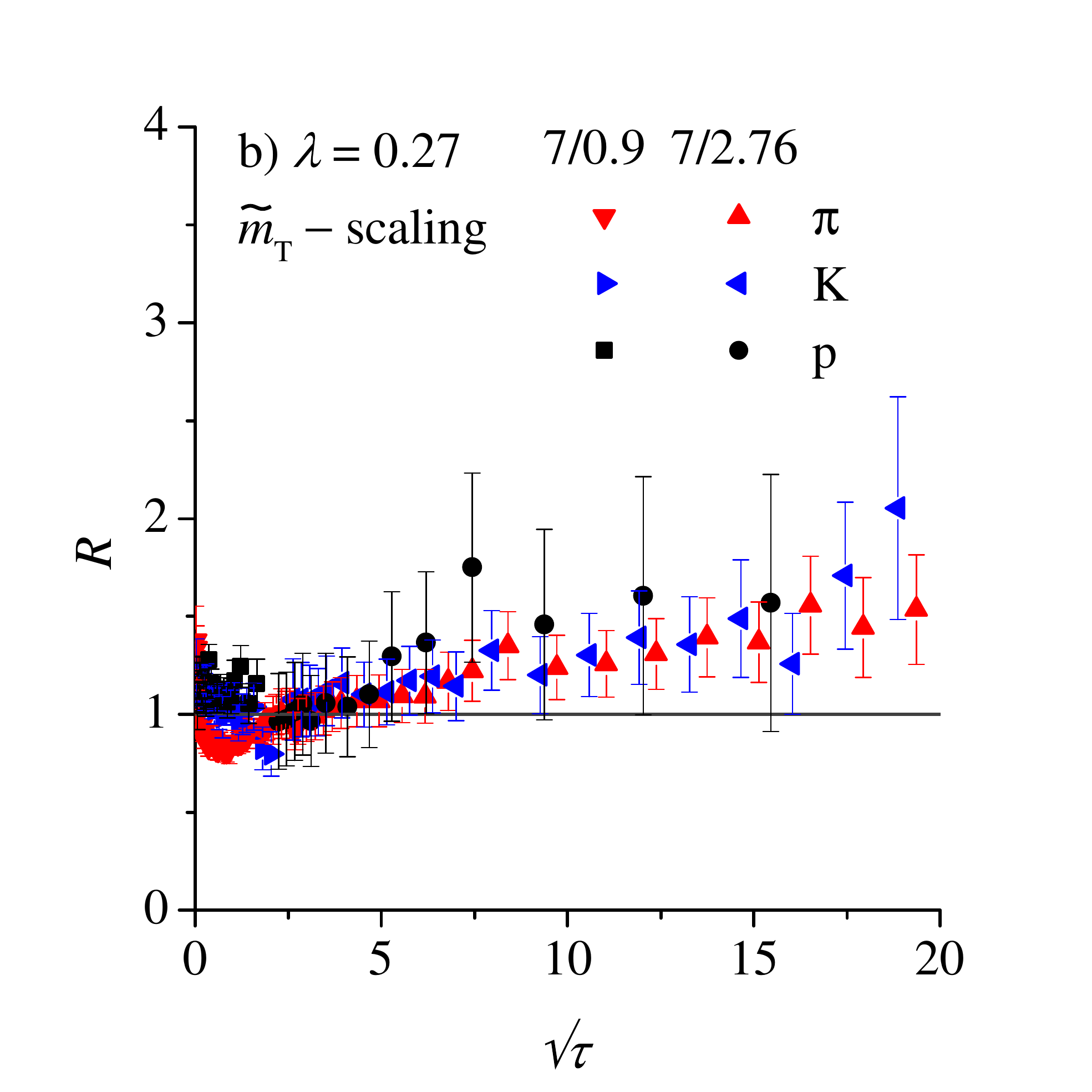}
\caption{Ratios $R$ from
Fig.~1.c and Fig.~\ref{ratiosmT}.c for extended horizontal axis.}%
\label{largeratios}%
\end{figure}

Quantitative analysis should also determine the $p_{\mathrm{T}}$ window where
GS should work. Here in Fig.~\ref{largeratios} we simply extend the $x$ axis
of Fig.~1.c and Fig.~\ref{ratiosmT}.c for the case of scaling in variable
$\tau_{p_{\rm T}}$ and $\tau_{\tilde{m}_{\mathrm{T}}}$ respectively. We see, as
expected from the properties of $\tilde{m}_{\mathrm{T}}(p_{\mathrm{T}})$ as
a function of transverse momentum, that the difference between the quality of GS
in these two variables shows only for small $\tau$'s. We can also see from
Fig.~\ref{largeratios} that GS window closes for $\tau\sim5$.

\begin{figure}[h!]
\centering
\includegraphics[scale=0.33]{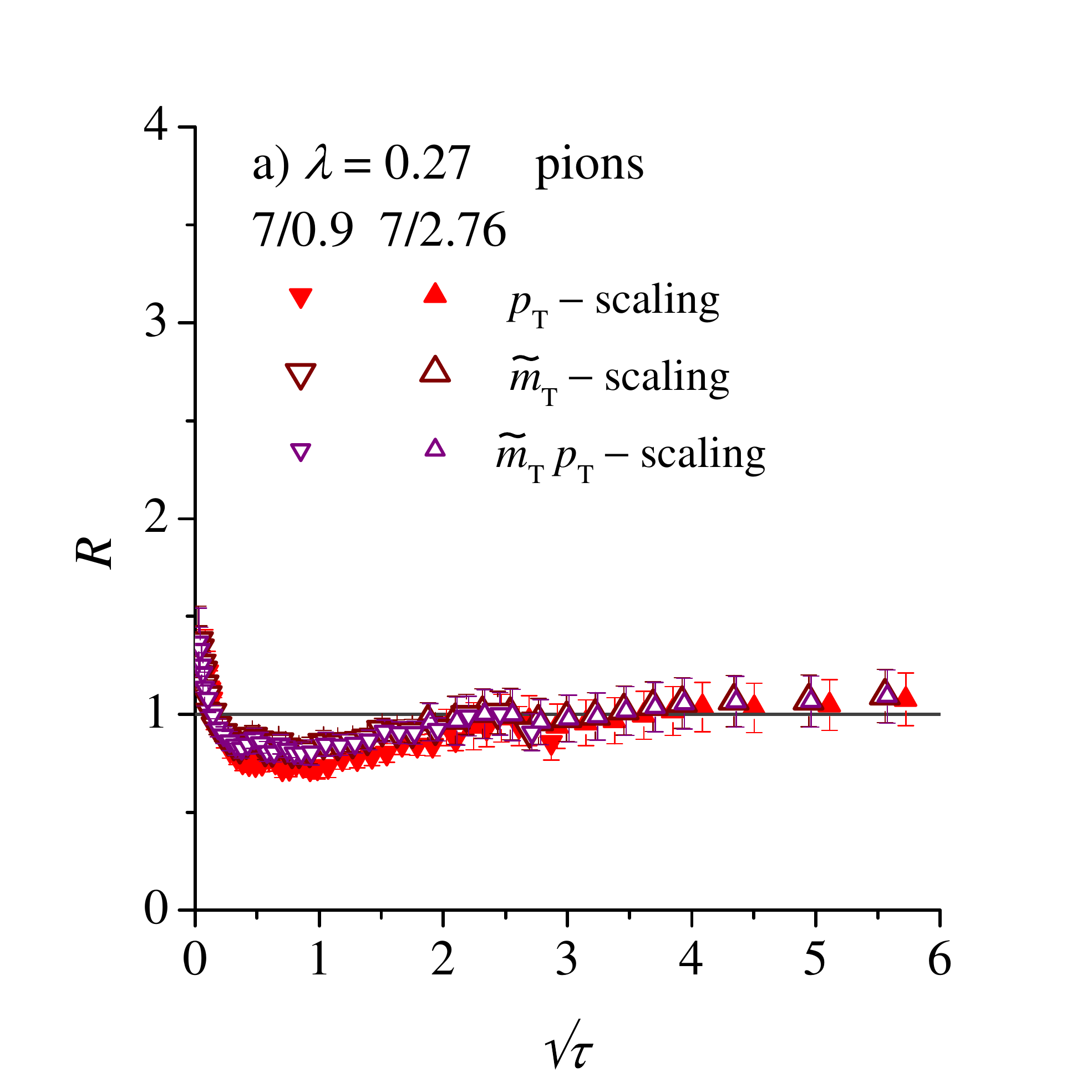}
\includegraphics[scale=0.33]{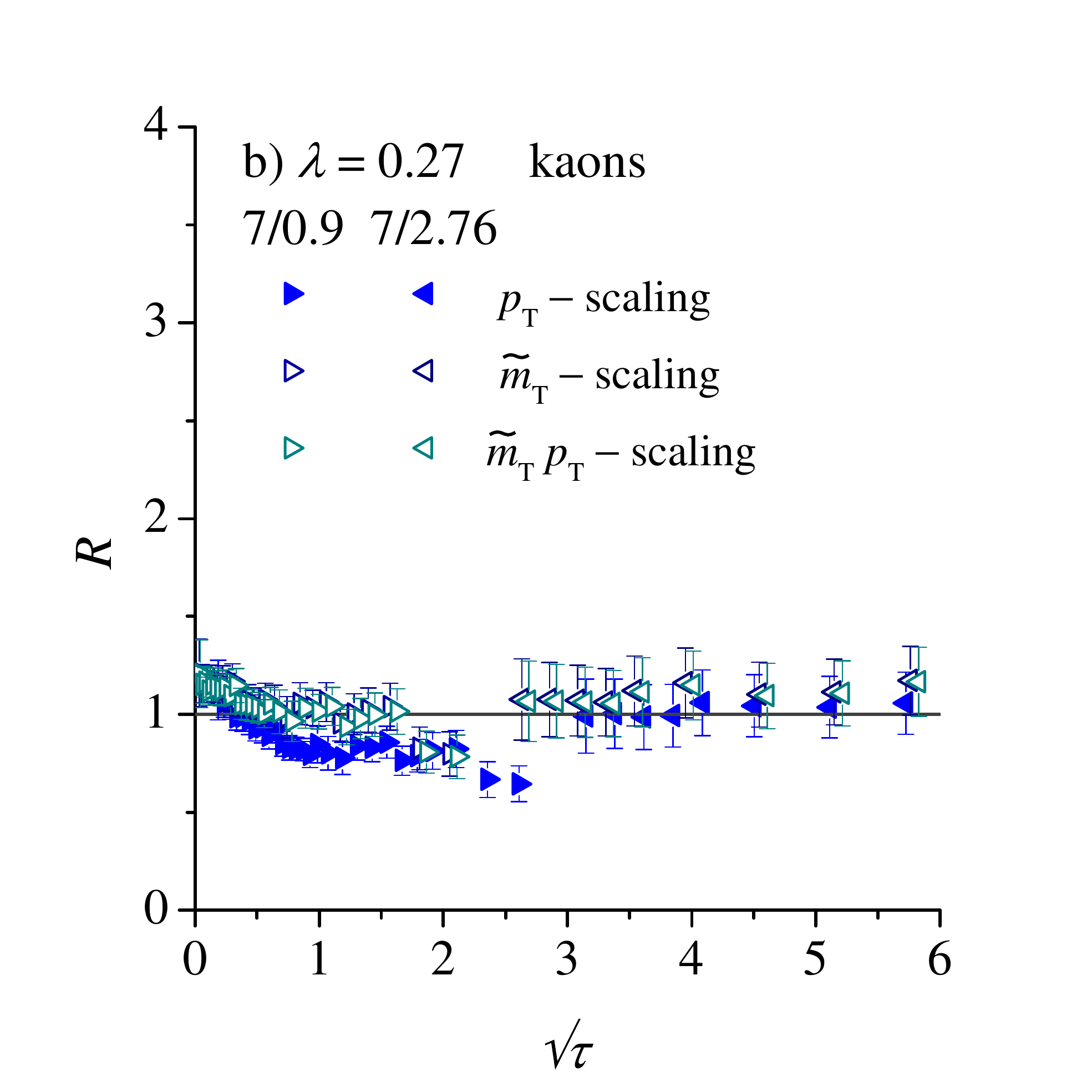}\\
\includegraphics[scale=0.33]{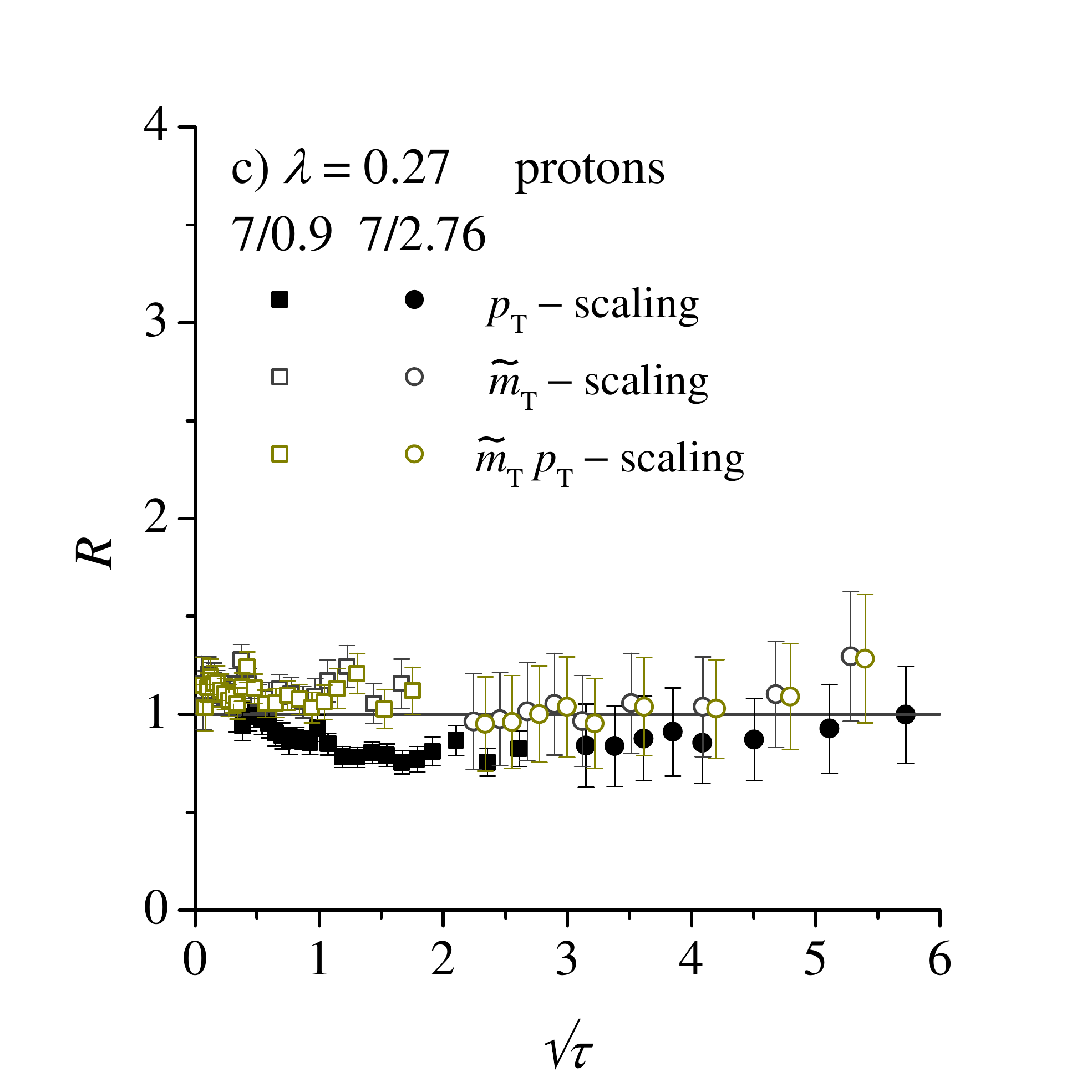}
\includegraphics[scale=0.33]{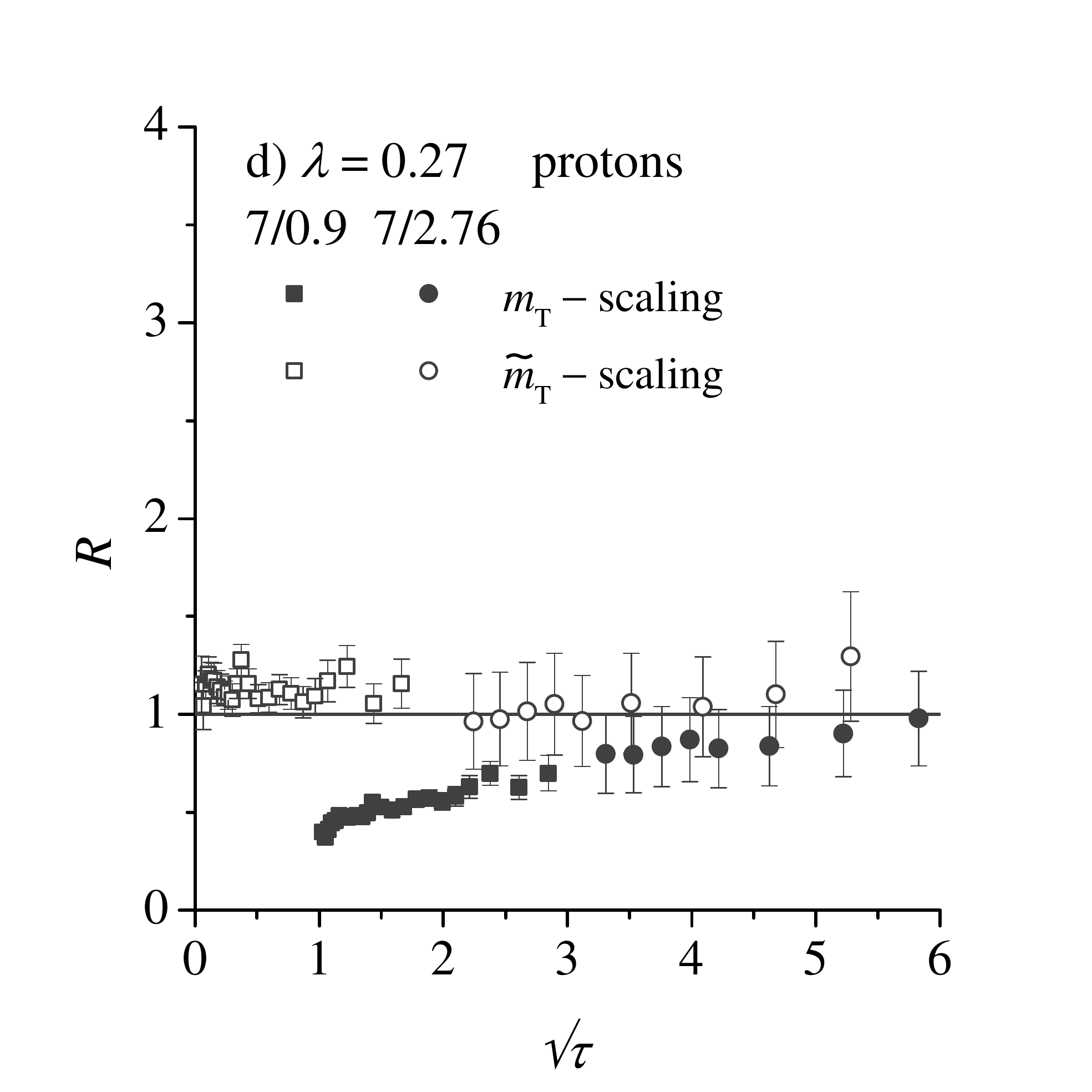}
\caption{Panels a) -- c):
comparison of geometrical scaling in three different variables: $\tau_{p_{\rm T}}$,
$\tau_{\tilde{m}_{\rm T}}$ and $\tau_{\tilde{m}_{\rm T} p_{\rm T}}$ 
for $\lambda=0.27$. Full symbols
correspond to ratios $R_{W_{1}/W_{2}}$ plotted in terms of the scaling
variable $\tau_{p_{\rm T}}$, open symbols to $\tau_{\tilde{m}_{\rm T}}$ 
and $\tau_{\tilde{m}_{\rm T}%
p_{\rm T}}$, note negligible differences between the latter two forms of scaling
variable. Panel a) corresponds to pions, b) to kaons and c) to protons. In
panel d) we show comparison of geometrical scaling for protons in scaling
variables $\tau_{\tilde{m}_{\rm T}}$ and $\tau_{m_{\rm T}}$, 
no GS can be achieved in the
latter case.}%
\label{ratiosmandpT}%
\end{figure}

Before closing this Section, let us see how scaling properties are affected by
going from scaling variable $\tau_{p_{\rm T}}$ (\ref{taupdef}) to $\tau_{\tilde
{m}_{\mathrm{T}}}$ (\ref{taumtdef}) and what would be the difference in
scaling properties if we had chosen $p_{\mathrm{T}}$ as an argument in the
saturation scale leading to scaling variable $\tau_{\tilde{m}_{\mathrm{T}%
}p_{\mathrm{T}}}$ (\ref{taumtpdef}), so called $\tilde{m}_{\mathrm{T}}%
$$p_{\mathrm{T}}$ -- scaling. This is illustrated in Fig.~\ref{ratiosmandpT}.a
-- \ref{ratiosmandpT}.c where full symbols refer to the $p_{\mathrm{T}}$ --
scaling (\ref{taupdef}) and open symbols to $\tilde{m}_{\mathrm{T}}$ --
scaling or $\tilde{m}_{\mathrm{T}}$$p_{\mathrm{T}}$ -- scaling. One can see
very small difference between open symbols indicating that scaling variables
$\tau_{\tilde{m}_{\mathrm{T}}}$ (\ref{taumtdef}) and $\tau_{\tilde
{m}_{\mathrm{T}}p_{\mathrm{T}}}$ (\ref{taumtpdef}) exhibit GS of the same
quality. On the contrary $p_{\mathrm{T}}$ -- scaling in variable $\tau_{p_{\rm T}}$
(\ref{taupdef}) is visibly worse than any form of scaling variable involving
$\tilde{m}_{\mathrm{T}}$.

Finally in Fig.~\ref{ratiosmandpT}.d, on the example of protons, we compare
$\tilde{m}_{\mathrm{T}}$ -- scaling and ${m}_{\mathrm{T}}$ -- scaling for
$\lambda=0.27$. One can see that no GS has been achieved in the latter case.
Qualitatively the same behavior can be observed for other values of $\lambda$.

Let us remark that recently CMS collaboration
has published data on identified spectra  \cite{Chatrchyan:2012qb}, however
for much smaller range of transverse momenta. Pion and proton spectra have
been measured up to 1.5 -- 1.7 GeV/$c$ respectively, whereas kaons
up to 1 GeV/$c$ only. In this region ratios $R_{W_1/W_2}$ develop 
a dip and therefore an attempt to draw conclusions on GS in this case may lead
to an underestimate of exponent $\lambda$ \cite{sikler2}.

The examples presented in this Section illustrate that for identified
particles geometrical scaling of good quality is observed for scaling variable
(\ref{taumtdef}) within the window $0.5 <\sqrt{\tau_{\tilde{m}_{\rm T}}} < 6$ for
kaons and protons with lower bound shifted to 1.5 for pions. In the next
Section we are going to investigate the consequences of this observation as
far as the universal shape of scaling function $F_{\alpha}(\tau)$ is concerned.

%%%%%%%%%%%%%%%%%%%%%%%%%%%%%%%%%%%%

\section{Consequences of geometrical scaling}

\label{conseq}

In what follows we shall assume that scaling variable $\tau=\tau_{\tilde
{m}_{\mathrm{T}}}$. We shall also suppress for the moment index $\alpha$ 
referring to the
particle species. Let us first examine the energy dependence of mid rapidity
multiplicity density and of average transverse momentum of produced particles.
Following Eq.~(\ref{GS}), mid rapidity density is given by an integral
\begin{equation}
\frac{dN}{dy}=\frac{1}{2\,Q_{0}^{2}}\int F(\tau) \,dp^{2}_{\text{T}%
}\label{GSint}%
\end{equation}
which requires change of variables:%
\begin{equation}
dp_{\text{T}}^{2}=2(\tilde{m}_{\text{T}}+m_{\alpha})\,d\tilde{m}_{\text{T}}%
\end{equation}
with%
\begin{align}
d\tilde{m}_{\text{T}}  & =\frac{Q_{0}}{2+\lambda}\left(  \frac{W}{Q_{0}%
}\right)  ^{\lambda/(2+\lambda)}\tau^{1/(2+\lambda)}\frac{d\tau}{\tau
}.\label{change}%
\end{align}
Using (\ref{change}) we arrive at (restoring dependence on particle species
$\alpha$)
%\begin{eqnarray}
%\frac{dN}{dy}
%&=&\frac{2}{2+\lambda }\left( \frac{W}{Q_{0}}\right) ^{\lambda /(2+\lambda
%)}\dint F_{\alpha }(\tau )\left( \left( \frac{W}{Q_{0}}\right) ^{\lambda
%/(2+\lambda )}\tau ^{1/(2+\lambda )}+\frac{m}{Q_{0}}\right) \,\tau
%^{1/(2+\lambda )}\frac{d\tau }{\tau }  \notag \\
%&=&\frac{2}{2+\lambda }\left[ \left( \frac{W}{Q_{0}}\right) ^{2\lambda
%/(2+\lambda )}\dint F(\tau )\tau ^{-\lambda /(2+\lambda )}d\tau
%\right.  \notag \\
%&&+\left. \frac{m}{Q_{0}}\left( \frac{W}{Q_{0}}\right) ^{\lambda
%/(2+\lambda )}\dint F(\tau )\tau ^{-(\lambda +1)/(2+\lambda
%)}d\tau \right] .
%\end{eqnarray}%
\begin{align}
\frac{dN_{\alpha}}{dy}  & =b_{\alpha}\left(  \frac{W}{Q_{0}}\right)
^{2\lambda/(2+\lambda)}\left[  1+\frac{a_{\alpha}}{b_{\alpha}}\frac{m_{\alpha
}}{Q_{0}}\left(  \frac{W}{Q_{0}}\right)  ^{-\lambda/(2+\lambda)}\right]
.\label{dNdy}%
\end{align}
We see therefore that mid rapidity identified particle density contains a
universal leading term and a correction proportional to the particle mass both
rising as powers of energy. The power like rise of mid rapidity density 
for charged (unidentified) particles has
been confirmed up to the LHC energies \cite{:2009dt}
and the leading power being 0.23 is in
agreement with $2\lambda/(2+\lambda) \approx0.23$ for $\lambda=0.27$
\cite{McLerran:2010ex} . For
large energies and small particle masses one can neglect the second term in
Eq.~(\ref{dNdy}). 

Constants $a_{\alpha}$ and $b_{\alpha}$ read:%
\begin{equation}
b_{\alpha}=\frac{1}{2+\lambda}%
%TCIMACRO{\dint }%
%BeginExpansion
{\displaystyle\int}
%EndExpansion
F_{\alpha}(\tau)\tau^{-\lambda/(2+\lambda)}d\tau,\;a_{\alpha}=\frac
{1}{2+\lambda}%
%TCIMACRO{\dint }%
%BeginExpansion
{\displaystyle\int}
%EndExpansion
F_{\alpha}(\tau)\tau^{-(\lambda+1)/(2+\lambda)}d\tau.\label{abalpha}%
\end{equation}

We shall show now  that GS leads  also to the power-like dependence of the
mean transverse momentum on the scattering energy. For massive particles we
have:%
\begin{align}
p_{\text{T}}  & =Q_{0}\left(  \frac{W}{Q_{0}}\right)  ^{\lambda/(2+\lambda
)}\tau^{1/(2+\lambda)}\sqrt{1+2\frac{m_{\alpha}}{Q_{0}}\left(  \frac{W}{Q_{0}%
}\right)  ^{-\lambda/(2+\lambda)}\tau^{-1/(2+\lambda)}}.
\end{align}
We see that for large energies the second term under the square root is
suppressed (and also for small masses) so after expansion for large $W$ we
obtain%
\begin{align}
p_{\text{T}}  & =Q_{0}\left(  \frac{W}{Q_{0}}\right)  ^{\lambda/(2+\lambda
)}\tau^{1/(2+\lambda)}+m_{\alpha}+ \ldots.
\end{align}

We define mean transverse momentum as:%
\begin{equation}
\left\langle p_{\text{T}}\right\rangle =\frac{\frac{1}{2 Q_{0}^{2}}
%TCIMACRO{\dint }%
%BeginExpansion
{\displaystyle\int}
%EndExpansion
p_{\text{T}}F_{\alpha}(\tau)dp_{\text{T}}^{2}}{\frac{1}{2 Q_{0}^{2}}
%TCIMACRO{\dint }%
%BeginExpansion
{\displaystyle\int}
%EndExpansion
F_{\alpha}(\tau)dp_{\text{T}}^{2}}.\label{meanpt}%
\end{equation}
Denominator of (\ref{meanpt}) is given by Eq.\thinspace(\ref{dNdy})
 whereas the numerator, after expanding in powers
of $m_{\alpha}$ reads
%\begin{equation}
%\text{num.}=\frac{2Q_{0}}{2+\lambda }\left( \frac{W}{Q_{0}}\right) ^{\lambda
%/(2+\lambda )}\dint \frac{d\tau }{\tau }\tau ^{1/(2+\lambda )}F_{\alpha
%}(\tau )\left( \left( \frac{W}{Q_{0}}\right) ^{\lambda /(2+\lambda )}\tau
%^{1/(2+\lambda )}+\frac{m_{\alpha }}{Q_{0}}\right) ^{2}.
%\end{equation}%
%Working with accuracy up to terms linear in $m/Q_{0}$ we get:%
\begin{align*}
\frac{\text{num.}}{Q_{0}}  & =\left(  \frac{W}{Q_{0}}\right)  ^{2\lambda
/(2+\lambda)}\left(  c_{\alpha}\left(  \frac{W}{Q_{0}}\right)  ^{\lambda
/(2+\lambda)}+2b_{\alpha}\frac{m_{\alpha}}{Q_{0}}\right)
\end{align*}
with%
\begin{equation}
c_{a}=\frac{1}{2+\lambda}%
%TCIMACRO{\dint }%
%BeginExpansion
{\displaystyle\int}
%EndExpansion
F_{\alpha}(\tau)\tau^{-(\lambda-1)/(2+\lambda)}d\tau.\label{calpha}%
\end{equation}
Hence mean $p_{\text{T}}$ reads%
\begin{equation}
\left\langle p_{\text{T}}\right\rangle =Q_{0}\frac{c_{\alpha}\left(  \frac
{W}{Q_{0}}\right)  ^{\lambda/(2+\lambda)}+2b_{\alpha}\frac{m_{\alpha}}{Q_{0}}%
}{b_{\alpha}+a_{\alpha}\frac{m_{\alpha}}{Q_{0}}\left(  \frac{W}{Q_{0}}\right)
^{-\lambda/(2+\lambda)}}\simeq Q_{0}\frac{c_{\alpha}}{b_{\alpha}}\left(
\frac{W}{Q_{0}}\right)  ^{\lambda/(2+\lambda)}+m_{\alpha}\left(
2-\frac{a_{\alpha}c_{\alpha}}{b_{\alpha}^{2}}\right)  .\label{meanpt1}%
\end{equation}
We see that mean transverse momentum behaves as a constant (proportional to
the particle mass) plus a power of energy, which is also confirmed by the recent
data up to the LHC energies \cite{:2009dt}. Let us remark that formulae (\ref{dNdy}) and
(\ref{meanpt1}) imply in the leading order
\begin{equation}
\left\langle p_{\text{T}}\right\rangle =A+B\sqrt{dN/dy}.
\end{equation}

%%%%%%%%%%%%%%%%%%%%%%%%%%%%%%%%%%%%%%

\section{Universal shape of geometrical scaling and Tsallis-like parametri{\-}zation}

\label{shape}

We shall now be more specific and use a particular form of function $F_{\alpha}(\tau)$.
This will allow us to calculate explicitly constants $a_{\alpha
},b_{\beta}$ and $c_{\alpha}$.
To this end we shall use the experimental observation that identified particles
spectra can be well described in terms of Tsallis-like parametrization of
Eq.~(\ref{Tsallis}) with species dependent temperature $T=T_{\alpha}$ and
exponent $n=n_{\alpha}$. In actual fits to the data $n_{\alpha}$ is of the
order $5$ to 9  \cite{Chatrchyan:2012qb}, therefore we may use an approximation%
\begin{equation}
C_{\alpha}\simeq\frac{\gamma_{\alpha} }{T_{\alpha}^{2}}\label{Cappr}%
\end{equation}
where constant $\gamma_{\alpha} $ restores the correct normalization being
only a function of $n_{\alpha}$. Inserting (\ref{Cappr}) and (\ref{change})
into (\ref{Tsallis}) we obtain:%
\begin{equation}
\frac{d^{2}N_{\alpha}}{dydp_{\text{T}}^{2}}=\frac{\gamma_{\alpha} }
{2T_{\alpha}^{2}}\frac{dN_{\alpha}}{dy}\left[  1+\frac{Q_{0}\left(  \frac
{W}{Q_{0}}\right)  ^{\lambda/(2+\lambda)}\tau^{1/(2+\lambda)}}{n_{\alpha
}\,T_{\alpha}}\right]  ^{-n_{\alpha}}.
\end{equation}
Finally, we shall use leading term for energy dependence of the mid rapidity
multiplicity distribution (\ref{dNdy}) which gives
\begin{equation}
\frac{d^{2}N_{\alpha}}{dydp_{\text{T}}^{2}}=\frac{\gamma_{\alpha} }
{2T_{\alpha}^{2}}b_{\alpha}\left(  \frac{W}{Q_{0}}\right)  ^{2\lambda
/(2+\lambda)}\left[  1+\frac{Q_{0}\left(  \frac{W}{Q_{0}}\right)
^{\lambda/(2+\lambda)}\tau^{1/(2+\lambda)}}{n_{\alpha}\,T_{\alpha}}\right]
^{-n_{\alpha}}.\label{dNetc}%
\end{equation}
The right hand side of Eq.(\ref{dNetc}) should be an energy independent
function of scaling variable $\tau$ only. Within approximations used so far
there exists a simple solution to this requirement:%
\begin{equation}
T_{\alpha}=\kappa_{\alpha} \,Q_{0}\left(  \frac{W}{Q_{0}}\right)
^{\lambda/(2+\lambda)}\label{Tsol}%
\end{equation}
where $\kappa_{\alpha}$ is a constant. Therefore GS predictions for Tsallis
parameters are that $T_{\alpha}$ depends on energy as a power\footnote{
This dependence is, however, rather weak for large energies.}, whereas
$n_{\alpha}$ is a constant. 
A complete fit to high energy data for charged (unidentified)
particles from NA49 energies up to the
LHC \cite{Rybczynski:2012vj} shows rather small variation (of the order of 10\%) 
of parameter $q(W)$
related to $n$ from Eq.~(\ref{Tsallis}) in the following way:
\begin{equation}
n(W)=\frac{1}{q(W)-1}
\end{equation}
which, however, translates into rather strong energy dependence of $n(W)$,
especially for smaller energies 
where $q(W)$ is only slightly bigger than 1.  
Similar conclusion --  as far as the energy dependence of the Tsallis parameters
for identified
particles
is concerned --  has been found in Ref.~\cite{Cleymans:2013rfq}  with temperature 
hardly depending on energy. One should note, however, that the multiplicity
distribution used in Ref.~\cite{Cleymans:2013rfq} slightly differs from the
one of Eq.~(\ref{Tsallis}).

The solution with constant
$n_{\alpha}$ has a number of corrections which in the present approach
can be studied in a systematic way. Ignoring them for them the moment
we arrive at
%
%\begin{equation}
%\frac{d^{2}N_{\alpha }}{dydp_{\text{T}}^{2}}=\frac{\gamma _{\alpha }}{%
%Q_{0}^{2}}\frac{b_{\alpha }}{2\kappa _{\alpha }^{2}}\left[ 1+\frac{\tau
%^{1/(2+\lambda )}}{n_{\alpha }\,\kappa _{\alpha }}\right] ^{-n_{\alpha }}.
%\end{equation}%
%This means that
the universal scaling function which takes the following form:%
\begin{equation}
F_{\alpha}(\tau)=\frac{\gamma_{\alpha}b_{\alpha}}{2\kappa_{\alpha}^{2}}\left[
1+\frac{\tau^{1/(2+\lambda)}}{n_{\alpha}\,\kappa_{\alpha}}\right]
^{-n_{\alpha}}.\label{Falpha}%
\end{equation}

The solution for $T_{\alpha}$ given by Eq.(\ref{Tsol}) can be interpreted in
terms of the saturation scale, $Q_{\text{s}}$ (\ref{Qsdef}) which for
$\tilde{m}_{\mathrm{T}}$--scaling takes the following form:%
\begin{equation}
Q_{\text{s}}(\tilde{m}_{\mathrm{T}})=Q_{0}\left(  \frac{\tilde{m}_{\mathrm{T}%
}}{W}\right)  ^{-\lambda/2}.
\end{equation}
For quantities integrated over transverse momentum one introduces another
saturation scale, $\bar{Q}_{\text{s}}$ which has a meaning of an average
transverse momentum, or in this case average value of $\tilde{m}_{\mathrm{T}}%
$, and can be thought of as a solution of an equation \cite{Kharzeev:2004if}:%
\[
\bar{Q}_{\text{s}}=Q_{\text{s}}(\bar{Q}_{\text{s}})
\]
which gives%
\begin{equation}
\bar{Q}_{\text{s}}=Q_{0}\left(  \frac{W}{Q_{0}}\right)  ^{\lambda/(2+\lambda
)}. \label{aveQsat}
\end{equation}
We see therefore that parameter $T_{\alpha}$, Tsallis temperature (\ref{Tsol}),
is proportional to the average saturation scale $\bar{Q}_{\text{s}}$ with
proportionality constant $\kappa_{\alpha}$ which depends on particle species
$\alpha$. Constants $\kappa_{\alpha}$ have been fitted to thermal
distributions in Ref.~\cite{McLerran:2013oju} and they are of the order of 0.1.
Similar solution for the unidentified spectra has been discussed recently
in Ref.~\cite{Rybczynski:2012pn}.

Constants $a_{\alpha},b_{\alpha}$ and $c_{\alpha}$ can be calculated
analytically for $F_{\alpha}(\tau)$ given by Eq.(\ref{Falpha}):
%Indeed,changing variables%
%\begin{equation*}
%y=\tau ^{1/(2+\lambda )}\,\rightarrow \,\tau =y^{(2+\lambda )}\,\rightarrow
%d\tau =(2+\lambda )y^{(1+\lambda )}dy
%\end{equation*}%
%\begin{equation}
%I_{\beta }=\frac{2}{2+\lambda }\dint d\tau \,\tau ^{\beta }\left[ 1+\frac{%
%\tau ^{1/(2+\lambda )}}{n_{\alpha }\,\kappa _{\alpha }}\right] ^{-n_{\alpha
%}}=2\dint dy\,y^{(2+\lambda )\beta +(1+\lambda )}\left[ 1+\frac{y}{n_{\alpha
%}\,\kappa _{\alpha }}\right] ^{-n_{\alpha }}.
%\end{equation}%
%Rescaling variable $y$ by $n_{\alpha }\,\kappa _{\alpha }$ we get%
%\begin{eqnarray}
%I_{\beta } &=&2(n_{\alpha }\,\kappa _{\alpha })^{(2+\lambda )(\beta
%+1)}\dint d\eta \,\eta ^{\lambda +1+(2+\lambda )\beta }\left[ 1+\eta \right]
%^{-n_{\alpha }}  \notag \\
%&=&2(n_{\alpha }\,\kappa _{\alpha })^{(2+\lambda )(\beta +1)}B((2+\lambda
%)(\beta +1),n_{\alpha }-(2+\lambda )(\beta +1))
%\end{eqnarray}%
%where
%\begin{equation*}
%B(x,y)=\dint dt\,t^{x-1}(1+t)^{-\left( x+y\right) }
%\end{equation*}%
%is Euler beta function. Hence%
%\begin{equation*}
%x=(2+\lambda )(\beta +1),\;y=n_{\alpha }-(2+\lambda )(\beta +1)
%\end{equation*}%
%and we have:%
%\begin{equation*}
%\begin{array}{rrrrr}
%& \beta  & \beta +1 & x & y \\
%a_{\alpha } & -\frac{\lambda +1}{\lambda +2} & \frac{1}{\lambda +2} & 1 &
%n_{\alpha }-1 \\
%b_{\alpha } & -\frac{\lambda }{\lambda +2} & \frac{2}{\lambda +2} & 2 &
%n_{\alpha }-2 \\
%c_{\alpha } & -\frac{\lambda -1}{\lambda +2} & \frac{3}{\lambda +2} & 3 &
%n_{\alpha }-3%
%\end{array}%
%\end{equation*}%
%Therefore we get:%
\begin{align}
a_{\alpha}  & =\frac{\gamma_{\alpha}b_{\alpha}}{2 \kappa_{\alpha}^{2}}%
(n_{\alpha}\,\kappa_{\alpha})B(1,n_{\alpha}-1),\nonumber\\
b_{\alpha}  & =\frac{\gamma_{\alpha}b_{\alpha}}{ 2\kappa_{\alpha}^{2}}%
(n_{\alpha}\,\kappa_{\alpha})^{2}B(2,n_{\alpha}-2),\nonumber\\
c_{\alpha}  & =\frac{\gamma_{\alpha}b_{\alpha}}{2 \kappa_{\alpha}^{2}}%
(n_{\alpha}\,\kappa_{\alpha})^{3}B(3,n_{\alpha}-3)\label{abc}%
\end{align}
where $B(x,y)$ is Euler beta function. The second equation (\ref{abc}) should
be understood as a normalization condition for $\gamma_{\alpha}$:%
\begin{equation}
\gamma_{\alpha}=\frac{2}{n_{\alpha}^{2}\,B(2,n_{\alpha}-2)}%
\end{equation}
which is independent of $\kappa_{\alpha}$, as expected. With this
normalization we arrive at:
\begin{equation}
a_{\alpha}=\frac{b_{\alpha}}{\kappa_{\alpha}} \frac{n_{\alpha}-2}{n_{\alpha}%
},\;\;\;\; c_{\alpha}={b_{\alpha}}{\kappa_{\alpha}} \frac{2 n_{\alpha}%
}{n_{\alpha}-3}%
\end{equation}
with $b_{\alpha}$ being a free constant which can be fitted from the energy
dependence of the mid rapidity density (\ref{dNdy}). The
coefficient governing the constant piece in a formula for $\left\langle
p_{\text{T}}\right\rangle $ (\ref{meanpt1}) is given by:%
\begin{align}
2-\frac{a_{\alpha}c_{\alpha}}{b_{\alpha}^{2}}  & =-\frac{2}{n_{\alpha}%
-3}.\label{coeffpt}%
\end{align}
It is important to note that this coefficient is negative for values of $n_{\alpha}$
extracted from the data \cite{Chatrchyan:2012qb} and that the
constant piece in Eq.~(\ref{meanpt1}) is growing with particle mass. 
Note, however, that there might
exist a nonperturbative contribution to this term which is beyond control in
the present approach.

%%%%%%%%%%%%%%%%%%%%%%%%%%%%%%%%%%%%%%%%

\section{Conclusions}

\label{concl}

In this paper we have demonstrated using recent ALICE 
pp data  for identified particles at three LHC energies
\cite{ALICE} that transverse momentum spectra exhibit geometrical scaling in
variable $\tau_{\tilde{m}_{\rm T}}=
(\tilde{m}_{\rm T}/Q_0)^{2} (\tilde{m}_{\rm T}/ W)^{\lambda}$.
It is impossible at present to asses quantitatively the quality of this scaling, since
the data for $W=0.9$ and $2.76$~TeV published so far, do not overlap (or have very
small overlap) in $p_{\rm T}$. It can be, however, seen "by eye" that  the
$\tilde{m}_{\rm T}$-scaling works better for identified particles than the
"standard" $p_{\rm T}$-scaling. Moreover, the optimal value of the exponent $\lambda$
is definitely closer to the DIS value of 0.32 than in the case of the $p_{\rm T}$-scaling
for (unidentified) charged particles where it is equal to 0.27 \cite{McLerran:2010ex}. 
This statements can be
quantified using the {\em method of ratios} \cite{Praszalowicz:2012zh}
 once data in the full $p_{\rm T}$ range is published.
 
 One of the immediate consequences of geometrical scaling in variable $\tau_{\tilde{m}_{\rm T}}$
 is power-like growth of multiplicity and average transverse momentum with
 scattering energy $W$. We have shown in Refs.~\cite{McLerran:2010ex} that the values of the pertinent exponents for (unidentified) charge particles are 
 in agreement with experimental fits. In the present work we have shown that the 
 coefficients of this growth and possible constant terms
 are calculable in terms of the universal scaling function $F(\tau)$ (\ref{abalpha},\ref{calpha}).
 From Eqs.~(\ref{dNdy}) and (\ref{meanpt1}) one can  in principle determine
 constants $a_{\alpha}$, $b_{\alpha}$ and $c_{\alpha}$ once the pertinent data is available.
 
 We have also made an attempt to predict constants $a_{\alpha}$, $b_{\alpha}$ and $c_{\alpha}$
 assuming certain form of the scaling function $F(\tau)$. To this end we have used an
 experimental observation that identified particle spectra for small and intermediate
 values of $p_{\rm T}$ are well described by the Tsallis-like parametrization (\ref{Tsallis}). 
 This allowed us to relate Tsallis temperature to the energy dependent average saturation
 scale $\bar{Q}_{\rm s}$ (\ref{aveQsat}). Within approximation used in this paper Tsallis
 exponent $n$ remains energy independent. Corrections to this solution are in principle
 calculable in the present approach.
 
 Phenomenological findings of the present paper call for deeper theoretical understanding.
 The meaning of constant $\kappa_{\alpha}$ in Eq.~(\ref{Tsol}) and species dependence of
 exponent $n_{\alpha}$ are the most obvious examples. This may, however, require to
 construct  a nonperturbative fragmentation model which is beyond the scope of
 the present paper.

\section*{Acknowledgements}

The author would like to thank Larry McLerran for discussion and remarks
and  the ALICE Collaboration for an access to the
data on the $p_{\rm T}$ spectra.
This research has been supported by the Polish NCN grant 2011/01/B/{\\}ST2/00492.

%%%%%%%%%%%%%%%%%%%%%%%%%%%%%%%%%%%%%%%%%%%%%%%

\end{document}